\documentclass[a4paper,fleqn,usenatbib]{mnras}

\usepackage[T1]{fontenc}
\usepackage{ae,aecompl}
\usepackage{graphicx}	
\usepackage{amsmath}	
\usepackage{amssymb}	

\usepackage{natbib}
\usepackage{graphicx}

\title[Disk-jet coupling]{`Spectro-temporal' variabilities and possible physical mechanism for Jet ejections}

\author[Radhika D. et al.]{
Radhika D.,$^{1,2}$\thanks{E-mail:radhikad\_isac@yahoo.in }
A. Nandi,$^{1}$
Agrawal V. K.$^{1}$
and S. Seetha$^{3}$
\\
$^{1}$Space Astronomy Group, ISRO Satellite Centre, ISITE, Outer Ring Road, Near Karthik Nagar, Marathahalli, Bangalore, 560037, India\\
$^{2}$Department of Physics, University of Calicut, Malappuram (District), 673635, India\\
$^{3}$Space Science Office, ISRO Headquarters, Antariksh Bhavan, New BEL Road, Bangalore, 560231, India \\
}

\date{Accepted XXX. Received YYY; in original form ZZZ}
\pubyear{2015}

\begin{document}
\label{firstpage}
\pagerange{\pageref{firstpage}--\pageref{lastpage}}
\maketitle

\begin{abstract}
In this paper, we attempt to find out the `spectro-temporal' characteristics during the jet ejection, of several outbursting Galactic black hole sources based on RXTE-PCA/HEXTE data in the energy band of 2 - 100 keV. We present results of detailed analysis of these sources during the rising phase of their outburst, whenever simultaneous or near-simultaneous X-ray and Radio observations are `available'. We find that before the peak radio flare (transient jet) a few of the sources (in addition to those reported earlier) exhibit `local' softening within the soft intermediate state itself. Except the duration, all the properties of the `local' softening (QPO not observed, reduction in total rms, soft spectra) are observed to be similar to the canonical soft state. We find similar `local' softening for the recent outburst of V404 Cyg also based on SWIFT observations. Fast changes in the `spectro-temporal' properties during the `local' softening implies that it may not be occurring due to change in Keplerian accretion rate. We discuss these results in the framework of the magnetized two component advective flow model.
\end{abstract}

\begin{keywords}
accretion, accretion disks -- black hole physics -- X-rays: binaries -- ISM: jets and outflows
\end{keywords}

\section{Introduction}
\label{intro}
Outbursting black hole (BH) binaries exhibit distinct and various types of spectral and temporal 
variabilities, and undergo spectral state transitions \citep{RM06} in their Hardness-Intensity 
diagram (HID) during their outburst phases \citep{TB2010}. In general, outbursting BH sources show signature of evolution of low
frequency QPOs (LFQPOs) during initial rising phases (\citealt{BH1990}, \citealt{Belloni2002}, 
\citealt{Casella2004}, \citealt{RM06}, \citealt{skcDeb08}, \citealt{Nandi2012}, \citealt{RN2014}, \citealt{INM15}) 
of the outburst (mainly in hard and hard intermediate states). The evolution of LFQPOs saturates in 
soft intermediate state and completely disappears or have weak presence during the soft state of the outbursts 
(\citealt{Casella2004}, \citealt{TB2010}, \citealt{Motta2011}, \citealt{Nandi2012}). 

In such sources, the presence of steady jets (\citealt{F01}, \citealt{CF02}) and strong relativistic jets have been observed (\citealt{Brock2002}, \citealt{Corbel2003}, \citealt{Corbel2004},
\citealt{FBG04}, \citealt{MJ2012}) which are mostly detected in radio waveband. \citealt{FBG04} have presented a unified picture for the disk-jet coupling phenomena occurring 
during the outburst evolution of BH binaries, based on several observational results 
(\citealt{Corbel01}, \citealt{GFP03}). The model connects the evolution of jets along-with the different 
spectral states. The model states that steady/persistent jets (Lorentz factor, $\Gamma$ $<$ 2) occur during the hard state (hereafter
LHS) and the 
hard intermediate state (hereafter HIMS) of BH sources. Jet ejections (Lorentz factor, $\Gamma$ $>$ 2) 
are observed during the transition from hard intermediate to soft intermediate state (hereafter 
SIMS), while they are suppressed during the soft state (hereafter HSS) (see also 
\citealt{FHB09} and \citealt{FB2012}). It has been also observed that during the transition from HIMS to SIMS, when a jet ejection 
occurs the type C QPO observed during HIMS disappears and a type A or B QPO is observed \citep{FBG04,Sol08}.  

\citet{Mirabel98} observed that when the inner disk disappeared, the X-ray 
brightness of the source GRS 1915$+$105 diminished, followed by increase in infrared and radio 
emission which indicated the formation of jets. It has been also observed that the 
sources XTE J1550$-$564 \citep{Corbel2002}, H 1743$-$322 \citep{Corbel2005} and XTE J1752$-$223 
\citep{Yang2010} showed the presence of large scale decelerating radio jets occurring 
after a long period of the main ejection, and it was found that these jets interact with the interstellar medium resulting in the acceleration of particles. 

Previous studies have shown that for the source GRS 1915$+$105 \citep{Feroci99, SV2001}, GX 339$-$4 \citep{FHB09} and 
H 1743$-$322 \citep{MJ2012}, when the transient relativistic jet ejection occurs, QPOs are 
not observed in the power spectra. Further \citealt{SV2001} observed that the Comptonized 
component also gets suppressed (see also \citealt{SV2003}), which was interpreted as 
due to ejection of matter from inner part (i.e., Comptonized corona) of the disk 
\citep{Nandi2001}. It has been observed that for the sources GRS 1915$+$105 and GRO J1655$-$40, 
whenever there is an ejection (i.e., loss of mass) from the inner part of the disk, the spectrum softens 
and during the process of an increase in mass, the spectrum gets harden \citep{Chak2002}. 
\citealt{FHB09} observed that during an ejection event (of a few BH sources), a drop in total 
variability (rms) occurs (also termed as low rms zones) in the power spectrum of the source along-with decrease in X-ray colour 
(hardness ratio).  

In the present work, we extend these previous studies to understand if the above 
characteristics are general, by investigating the `spectro-temporal' variations for several 
outbursting BH sources (in addition to those already observed) before the 
radio peak of the transient jet. We attempt to understand the phenomenon based on the possible physical mechanisms, especially within 
the framework of the 
two component advective flow model in the presence of magnetic field. \citealt{Nandi2001} has 
explained briefly how the changes occurring within the disk due to amplification of magnetic field, and their collapse can result into formation of jets, and these features can be evident in the `spectro-temporal' variations indicated by QPOs not being observed 
and spectral softening (see \citealt{SV2001}). Hence this paper is based on the context of \citealt{Nandi2001} and \citealt{SV2001}, and we study the 
characteristics of QPOs, power spectra and energy spectra before, during and after the triggering of jet ejection or the peak of the radio flare 
to understand the disk-jet coupling. We re-investigate the `spectro-temporal' characteristics of different outbursting black hole sources during the jet ejections. We also look into the timescale taken for these events, and understand the possible reasons for the same based on accretion flows corresponding to different timescales. 

We consider the observations at the time of {\it simultaneous or quasi-simultaneous} X-ray and Radio observations of nine BH sources namely GX 339$-$4, XTE J1859$+$226, V404 Cyg, H 1743$-$322, XTE J1752$-$223, GRO J1655$-$40, XTE J1748$-$288, XTE J1550$-$564 and MAXI J1836$-$194. We also attempt to find out the 
possible connection between the duration of SIMS and HSS, w.r.t the ejection event and to estimate the timescale for these events. 
 All these are achieved based on, spectral and temporal analysis of the observations performed by RXTE - PCA and HEXTE (in 2 - 100 keV energy band), SWIFT - XRT and BAT in 0.5 - 100 keV energy band and, the observations by different Radio observatories. The variations in the ASM hardness ratio are also studied during the period of ejection. We understand that a few of the sources (GX 339$-$4, XTE J1859$+$226, V404 Cyg and H 1743$-$322), exhibit a `local' softening before the detection of a peak radio flare, in addition to the characteristics already reported. A few other sources (XTE J1752$-$223, GRO J1655$-$40, XTE J1748$-$288 and XTE J1550$-$564) show a canonical softening. An estimate of the timescale of the rising phase is performed, so as to find out its connection with amplification of magnetic field. We also look into the properties of the companion stars of the black hole binaries studied, in order to verify whether they are magnetically active or not.

This paper is organized as follows. In the next section, we summarize the details of observations 
considered and the methodology applied for analyzing the archival data. In \S \ref{results}, the 
results of X-ray characteristics (i.e., temporal and spectral features) observed for 
the sources during the radio ejections are presented. A summary of the different possible physical mechanisms for 
the disk-jet coupling phenomenon are described in \S \ref{models}. We discuss these results in detail in \S \ref{Discu} with an attempt to interpret the disk-jet coupling observed in 
the outbursting BH sources. In \S \ref{Conc} we summarize the conclusions from our study.

\section{Observations and data analysis}
\label{obsanal}
We re-analyze the public archival data obtained from the HEASARC database for the RXTE mission during 
the period from 1998 to 2011 in order to study the evolution of temporal and spectral features of the BH 
sources during the radio flares. The standard procedures for PCA and HEXTE data reduction is employed 
using the FTOOLS package HEASOFT v 6.15.1. 

For timing analysis, we use PCA science data of Binned mode, Event mode and Good Xenon data. 
Light curves are generated in the energy bands of 2 - 6 keV, 6 - 13 keV, 13 - 25 keV and 
2 - 25 keV in order to perform an energy dependent study of the Power Density Spectrum (PDS). 
Energy dependent study of the PDS is performed using the customized IDL based software `General High energy Aperiodic Timing Software (GHATS)' 
\footnote{http://www.brera.inaf.it/utenti/belloni/GHATS\_Package/Home.html}. 
The GHATS package considers dead time effect and hence applies the correct normalization 
factor (as in \citealt{Zhang1995}) while subtracting the white noise during the generation of the 
PDS. We choose the minimum time resolution of 0.00781 sec corresponding to Nyquist frequency of 64 Hz, and generate power spectra for 8192 bins (64 sec). From this the averaged power spectra are created for every observation, and a re-binning factor of -1.01 is applied. The power obtained 
in the PDS has units of rms$^2$/Hz. The PDS is modeled from 0.1 to 64 Hz using different 
components of Lorentzians and power-law for any red noise observed. We estimate the centroid
QPO frequency, the Q-factor, significance and amplitude of the QPO, and the integrated rms for 
the overall PDS in the range of 0.1 to 64 Hz following the methodology in 
\citealt{Casella2004}, \citealt{RN2014}. 
The error values on each parameter of the components are estimated at 90\% confidence interval. The 
classification of QPOs into types 
of A, B, C or C* have been taken into account based on \citealt{Wiji99}, \citealt{Casella2004}, \citealt{Motta2011}.

Spectral data are extracted using Standard2 data product when only PCU2 is ON since it has the best efficiency. We follow the standard procedures of data reduction mentioned in RXTE Cookbook 
in order to generate the PCA and HEXTE spectra along-with their spectral responses 
(see also \citealt{RN2014} for details of data reduction). 
Since the HEXTE cluster B was left in off source position since December 2009, the procedures to 
be applied for data reduction for observations after this period, is different from the above. 
We refer to the HEXTE news release 
\footnote{http://heasarc.gsfc.nasa.gov/docs/xte/whatsnew/\\newsarchive\_2010.html\#hexteB\_locked} 
for this and also the suggestions provided by the HEXTE team (private communication), in order to extract the 
scientific spectral data. Using \emph{saextrct} we initially generate the spectrum for cluster A source and cluster B background data. 
The ftool \emph{hextebackest} is used to generate background file of 
cluster A on-source data, from cluster B off-source data. Since the resultant spectrum from array 
mode data shows broad peaks at 30 keV and 70 keV, upon the suggestion of the HEXTE team we extract the spectrum from event mode data which has finer binning than array data, using \emph{seextrct}. Once the source and 
background spectral files are obtained using \emph{seextrct} and \emph{hextebackest}, after applying dead time corrections, response files 
are generated using \emph{hxtrsp}. The resultant spectrum does not 
show any broad features. We apply this method for generating spectra and responses from HEXTE 
data for the sources XTE J1752$-$223 and MAXI J1836$-$194 where the observations considered are after 2009.
 
The package XSPEC v 12.8.1 is used for the purpose of spectral analysis. Simultaneous fitting of 
the PCA \& HEXTE data are carried out over the energy range of 3 - 100 keV. The spectra are 
modeled using a thermal {\it diskbb} \citep{Mitsuda84} component, a non-thermal component ({\it powerlaw} or {\it powerlaw 
+ highecut} of XSPEC) modified by the interstellar absorption (see \citealt{RN2014} for details). 
We estimate the total flux in the 3 - 100 keV energy band, with the help of the \textit{flux} 
command along-with its error at 90\% confidence interval. The PCA hardness ratio is estimated as ratio of counts in 6 - 20 keV and 2 - 6 keV energy bands. 
In order to understand the evolution of 
spectral components (thermal and non-thermal) during a jet ejection event, we also estimate the 
fractional contribution of disk flux and powerlaw flux in the energy range of 3 - 100 keV with the help of the convolution model \textit{cflux}.
 The fractional
contributions of disk (thermal component) and powerlaw flux (non-thermal component) over total flux 
have been represented in Figure \ref{ratio-time} for all the sources (except MAXI J1836$-$194) studied in the present work. 

The one day averaged data of ASM in its different energy bands have been obtained. An estimate of the hardness ratio is performed by taking ratio of flux in the bands 5 - 12 keV and 3 - 5 keV, which are represented in Figure \ref{asm}, Tables \ref{flares}, \ref{flares2} and \ref{flares3}, along-with other `spectro-temporal' parameters.

Simultaneous or quasi-simultaneous observations carried out by the several Radio observatories like 
VLA, MERLIN, RYLE, ATCA, RATAN etc., are considered in order to understand the evolution of radio 
flux during the outburst of the BH sources (\citealt{Hjel6934,Hanni2001,Brock2002,Gallo2004,Rupen2005,
Shapo2007,Trushkin2011,MJ2011,MJ2012,Brock13} and see references therein).

In addition to these, we have also looked into the SWIFT observations of V404 Cyg (GS 2023$+$338) during its June 2015 outburst. The methodology employed for the `spectro-temporal' analysis of this source has been discussed in detail in \citealt{RAVS16}(arxiv:1601.03234). The radio observations performed by RATAN-600 \citep{2015ATel.7716....1T}, AMI-LA \citep{2015ATel.7658....1M,2015ATel.7714....1M} and VLA, SMA and JCMT SCUBA-2 \citep{2015ATel.7708....1T} are also referred to. 

\section{Results}
\label{results}
In this section, we present the results on the spectral and temporal variations during the rising phase of the outbursts whenever radio 
flares i.e., when the jet ejections have taken place, to understand the disk-jet coupling for 
several BH sources (as mentioned in \S 1). 

\subsection{Rapid transition within SIMS}
We find that a few of the BH sources exhibit rapid changes (i.e. in faster timescales) in their `spectro-temporal' characteristics, just after the triggering of the jet ejection/before the detection of the peak radio flare. In this sub-section we summarize the results for these sources.
\subsubsection{GX 339$-$4}

\begin{itemize}

\item 2002 outburst 

We observe type C QPOs during the LHS and HIMS of GX 339$-$4 during its 2002 outburst. A type C QPO of 5.77 Hz is observed on MJD 52406.70. The next X-ray observation is during MJD 52410.53 only (see also Table 3 of \citealt{Motta2011}). Hence due to lack of PCA and HEXTE observations between 
MJD 52406.70 and MJD 52410.53, the previous works (see \citealt{Belloni2005,Motta2011}) could not study the X-ray characteristics during this period. In this work, 
we find that the ASM hardness ratio decreases during this period (see Table \ref{flares}, arrow mark in Figure \ref{asm}), and a sudden increase in 
radio flux from 10 mJy to 55 mJy on MJD 52408.48 has been observed by \citealt{Gallo2004}. 

A type C QPO of 7.96 Hz is observed during MJD 52410.53, and as the source transits from HIMS to SIMS, on MJD 52411.60 we observe a type A QPO of 7.01 Hz, followed by a type B QPO of 
5.91 Hz on MJD 52411.65. Subsequent observations (MJD 52412.07 and 52414.35) do not 
show the presence of QPOs, and the total rms of the PDS also reduces to $\sim$ 2.5\% (see also \citealt{FHB09}). We also observe a decrease in 3 - 100 keV flux along-with increase in the disk flux contribution to the total flux (see Figure \ref{ratio-time}). We also find reduction in both the PCA and ASM hardness ratios (Table \ref{flares} and Figure \ref{asm}). A radio flare with flux of 20.39 mJy is observed on MJD 52413.42 \citep{Gallo2004}. We observe that the `spectro-temporal' characteristics during this short period of $>$ 2 days, are  similar to that observed in canonical HSS, and also are repetitive within the SIMS.   
  
\item 2010 outburst

We observe type C QPOs during the LHS and HIMS of GX 339$-$4 in its 2010 outburst. A type C QPO of 5.7 Hz is observed during MJD 55303.6, and when the source transits to SIMS on MJD 55304.71, a type B 
QPO of 5.71 Hz is observed. \citep{Motta2011,Nandi2012}. This SIMS exists for $>$1.5 days. The fractional contribution of disk flux increases from 0.45 to 0.67, while that of powerlaw flux 
reduces from 0.54 to 0.32 (Figure \ref{ratio-time}). The following observations for a period 
of $\sim$ 3 days (MJD 55305 to 55307), shows no QPOs in the PDS. During this period the 3 - 100 keV 
flux reduces from 11.6 $\times$10$^{-9}$ erg cm$^{-2}$ sec$^{-1}$ 
to 8.06 $\times$10$^{-9}$ erg cm$^{-2}$ sec$^{-1}$. The fractional contribution of disk flux 
increases to 0.8, while the powerlaw flux contribution is observed to decrease to 0.19
(Figure \ref{ratio-time}). The PCA hardness ratio decreases to 0.14, and the ASM hardness ratio 
reduces to 0.30 during this period. The total rms decreases to $\sim$ 3\% during MJD
 55306 and MJD 55307. Thus these three observations indicate softer spectral characteristics, in similarity with `canonical' HSS 
(shaded region in Figure \ref{fig1-GX-2010}). This is similar to the changes we have observed in the 2002 outburst of GX 339$-$4, and none of the previous studies have reported this.
 There have been reports of quenching of optical/IR flux during this 
period suggesting radio flares to have occurred during the period of MJD 55303 (17th April) to 
MJD 55308 (22nd April) \citep{Russell2010}, which has been later mentioned by \citealt{CadolleBel2011}. 
But the radio flux level has not been reported so far. 

A type B QPO of 5.55 Hz is observed on MJD 55308.98 (see also \citealt{Motta2011}). During this observation 
the total rms of the PDS increases to 6\% and the fractional disk flux contribution decreases to 0.32. PDS of 
observations (MJD 55310.03 on-wards) do not show QPOs and the 3 - 100 keV flux decreases to 
7.4 $\times$10$^{-9}$ erg cm$^{-2}$ sec$^{-1}$ indicating another softening similar to that observed earlier. Thus we find that the rapid change in the `spectro-temporal' characteristics observed is repetitive as the outburst progresses within the SIMS.

\begin{figure}
\centering
\includegraphics[width=9cm]{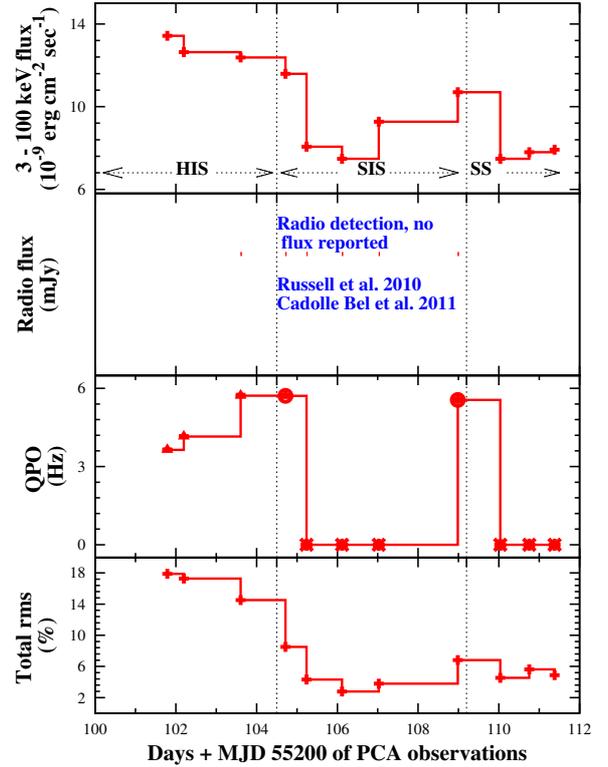}
\caption[`Spectro-temporal' evolution during 2010 outburst of GX 339$-$4]{Evolution of X-ray flux (3 - 100 keV), Radio flux, QPO frequency and total rms of the PDS in 
2 - 25 keV energy band for 2010 outburst of GX 339$-$4. The different spectral states are also marked. In the 3rd panel, we have represented type C QPOs with triangular points, type B with circular points and observations without QPOs using stars.}
\label{fig1-GX-2010}
\end{figure}

The repetitive softening observed in the 2002 and 2010 outburst within the SIMS has been verified by us \citep{RN2014} during the multiple flares of XTE J1859$+$226 and is described in brief in the following subsection.
\end{itemize}

\subsubsection{XTE J1859$+$226 (1999 outburst)}

\begin{figure}
\includegraphics[width=9cm]{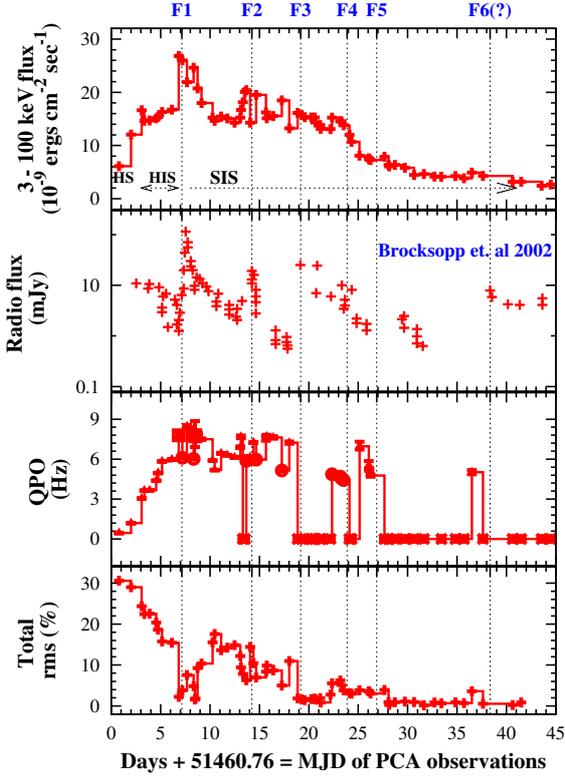}
\caption{Evolution of X-ray flux (3 - 100 keV), Radio flux, QPO frequency and total rms of the 
power spectra in 2 - 25 keV energy band during the 1999 outburst of XTE J1859$+$226. QPO not 
observed partially in 2 - 5 keV and 13 - 25 keV during F1; possible ejection of 6th flare (F6) 
is also marked (see \citealt{RN2014} for details). X-ray spectral states and time of peak 
radio flares are marked. In the 3rd panel of this figure, the type A QPOs are denoted using squares, type B using circles, type C using triangles and the stars represent those observations where QPOs are not observed.}
\label{fig1-1859}
\end{figure}

Type C QPOs (Figure \ref{fig1-1859}) are observed during 
the LHS and HIMS, and as the source enters into the SIMS, we observe a 
type A QPO followed by increase in radio flux which results 
in the first flare F1 (\citealt{Brock2002}, \citealt{FHB09}, \citealt{RN2014}). 
It has been observed that just after the ejection of the flares F2, F3, F4 and F5, QPOs are not observed in the 2 - 25 keV energy band, while during F1, QPOs are not observed in partial energy bands.  
Total rms is observed to decrease, along-with softening of the spectrum indicated by increase in fractional disk flux (Figure \ref{ratio-time})
 and also a drop in both the PCA and ASM hardness ratios (see Figure \ref{fig1-1859} and \ref{asm}, and also \citealt{RN2014} for details). The observations during which such characteristics are observed, varies over a timescale of 1 day to 8 days, and are found to be similar to that of the canonical HSS (see also Table \ref{flares} and \ref{flares2}). Details of `spectro-temporal' features during each of the radio flares are discussed in 
\citealt{RN2014}. 

\subsubsection{GS 2023$+$338/V404 Cyg (2015 outburst)}

Recently, we have also looked into the disk-jet coupling observed in the June 2015 outburst of the source GS 2023$+$338, which is well-known as V404 Cyg, based on its SWIFT observations. This source has been observed to exhibit multiple flaring activities throughout the electromagnetic spectrum. We have studied the detailed `spectro-temporal' behaviour of this source and explained in \citealt{RAVS16}(arxiv:1601.03234).

Three radio flares have been observed during the rising phase of the outburst of V404 Cyg. We find that as the source transits from the hard to the intermediate states, an oscillating radio flare of 140 mJy at 16 GHz has been detected on MJD 57192 \citep{2015ATel.7658....1M,2015ATel.7714....1M,2015ATel.7716....1T}. Around 12 hrs before this, we observe that both the hardness ratio and the fractional rms variability of the PDS declines and the X-ray flux in 0.5 - 10 keV increases. The second radio flare of 336.24 mJy at 8.2 GHz is reported on MJD 57195 \citep{2015ATel.7708....1T,2015ATel.7716....1T}, while the source exists in the intermediate state. Just before this flare, i.e. $\approx$ 10 hrs before, we find that 0.5 - 10 keV flux increases, while the 15 - 150 keV flux, the hardness ratio and rms variability of the PDS decreases. We find similar characteristics during the peak radio flare of 4085.1 mJy at 8.2 GHz on MJD 57199 \citep{2015ATel.7716....1T} when the source is in the intermediate state. A significant decrease is observed in the 15 - 150 keV hard flux, hardness ratio and the rms variability, on MJD 57198.01 which is $\approx$ 1 day before the peak of the radio flare. We also observe that during this observation the fractional contribution of disk flux increases (from 0.6 to 0.8) while that of the powerlaw flux decreases (0.4 to 0.2).

\subsubsection{H 1743$-$322 (2009 outburst)}

\begin{figure}
\includegraphics[width=9cm]{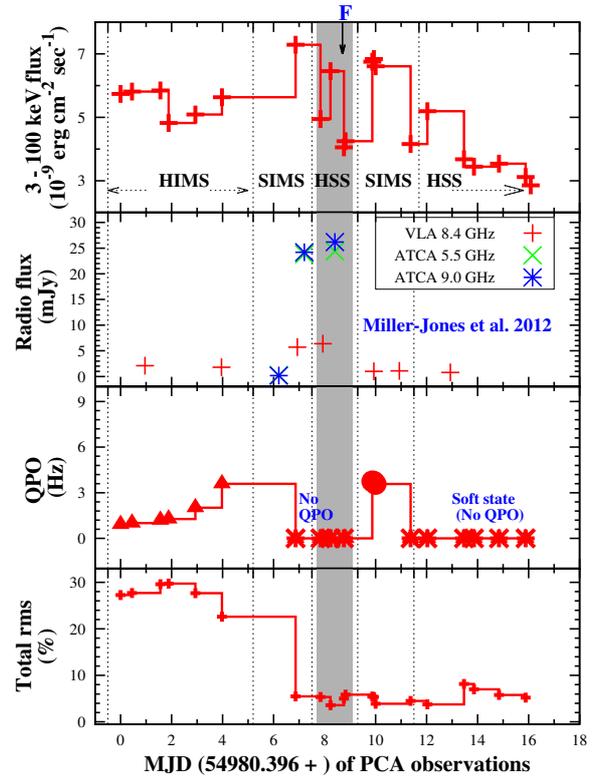}
\caption{Evolution of X-ray flux (3 - 100 keV), Radio flux, QPO frequency and total rms of the 
power spectra in 2 - 25 keV energy band during the 2009 outburst of H 1743$-$322; the X-ray 
spectral states and time of peak radio flare are marked. In the 3rd panel, type C QPOs are denoted using triangles, type B with circles and the stars corresponds to observations where QPOs are not observed.}
\label{fig1-H}
\end{figure}

Similar to that presented by \citealt{Motta2010,MJ2012}, we also observe that during the transition from HIMS to SIMS (which existed for a duration of $>$2 days), on MJD 54987.26, QPOs 
are not observed in the PDS (Figure \ref{fig1-H}), and the total rms of the PDS reduces. We observe that during this period the fractional contribution of the disk flux increases while that of the powerlaw flux decreases (Figure \ref{ratio-time}). Along-with these, we observe decline in the PCA and ASM hardness ratios (see Table \ref{flares2} and 
Figure \ref{asm}). Radio observations indicate 
the moment of jet ejection on MJD 54987.33 at 5.7 mJy at 8.4 GHz based on VLA observations \citep{MJ2012}, which is within 90 minutes after the non-detection of QPO on MJD 54987.26. QPOs are not observed on MJD 54988 and MJD 54989 during which we observe a softer spectra, PCA hardness ratio is found to have reduced to $\leq$ 0.4 and the total rms remains to be less around 3\% (see Figure \ref{fig1-H}, Table \ref{flares2}). \citealt{Motta2010} considers the source to have occupied a HSS during this period of $\sim$2 days. Although these properties are similar to that observed in the canonical HSS of this source, probability of this period being a canonical HSS or not is explained in the discussion section. 

\subsection{Canonical transition to HSS}
We observe that a few sources occupy the canonical HSS, after the jet ejection/before peak radio flare. This canonical HSS occurs for a longer duration, and in this sub-section we summarize the `spectro-temporal' characteristics of these sources during the jet ejections.

\subsubsection{XTE J1752$-$223 (2009 outburst)}

\begin{figure}
\centering
\includegraphics[width=9cm]{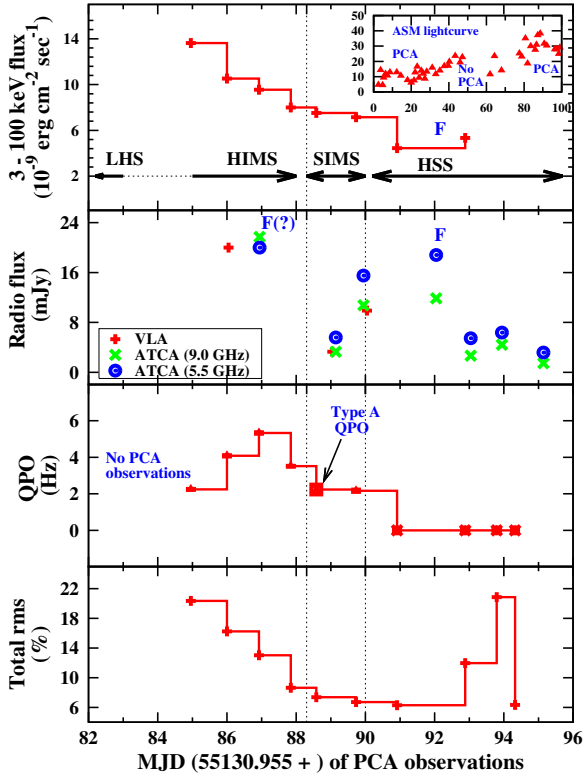}
\caption[`Spectro-temporal' evolution of XTE J1752$-$223]{Evolution of X-ray flux (3 - 100 keV), Radio flux, QPO frequency and total rms of the PDS during 2009-2010 outburst of XTE J1752$-$223. Here, in the 3rd panel we have used triangular points to denote type C QPOs, squares for type A and stars for observations without QPOs.}
\label{fig1-1752}
\end{figure}

As the source transits to the SIMS, on MJD 55219.53, a type A QPO of 2.23 Hz (Q-factor = 0.96, significance = 5.66) is observed (Figure \ref{fig1-1752}). We also find that this QPO is not observed in 2 - 5 keV and 13 - 25 keV energy bands. The total rms of the PDS is 
observed to reduce to 7.36\%, and the ASM hardness ratio also decreases (see Table \ref{flares3}). Although the 3 - 100 keV flux reduces to 7.51 $\times$10$^{-9}$ erg cm$^{-2}$ sec$^{-1}$, the fractional contribution of 
disk flux increases from 0.38 to 0.44, and that of powerlaw flux decreases from 0.61 to 0.56 (Figure \ref{ratio-time}).
Radio observations indicate the increase in flux from 6 mJy (at 5 GHz) on 
MJD 55220.1 to a peak radio flare of 18.8 mJy (at 5 GHz) on 
MJD 55223 \citep{Brock13}. QPOs of type C with frequency of 2.16 Hz are observed, 
during MJD 55220.68. From the next observation, QPOs are not observed and the total 
rms of the PDS decreases, and both the PCA and ASM hardness ratios also drops down as the source 
enters the canonical HSS which lasts for $\sim$60 days (see Table \ref{flares3}). Spectral softening is also indicated by an increase in disk flux in Figure \ref{ratio-time}.

\subsubsection{GRO J1655$-$40 (2005 outburst)}

We find that type C QPOs are observed during HIMS \citep{Shapo2007,Deb08,skcDeb08} and the transition 
to SIMS is marked by presence of a type B QPO of 6.7 Hz on MJD 53440.7, followed 
by a weak detection of a type C QPO of 16.4 Hz (significance of 3.1, Q-factor of 5.9) on MJD 53441.51. 
For observations since MJD 53441.59, QPOs are not observed in the PDS, 
the total rms is observed to decrease to 3.5\%. The fractional 
contribution of disk flux increases from 0.12 to 0.4 and that of powerlaw flux decreases from 0.88 to 0.6 (Figure \ref{ratio-time}). Both the PCA and ASM hardness ratios are observed to decrease (see Figure \ref{asm} and Table \ref{flares3}). The source moves towards the HSS \citep{Shapo2007, Motta2012} after the 
SIMS which lasted only for $\sim$2 days. There also 
have been report of weak detection of 1.6 mJy radio flux during this period,
 which is observed to attain a peak flux of 5.9 mJy on MJD 53449.8, around 8 days later 
(\citealt{Rupen2005}, \citealt{Shapo2007}). The canonical HSS occurs for a duration of $\sim$57 days.

\subsubsection{XTE J1748$-$288 (1998 outburst)}

During the rising phase, while the 
source existed in the very high state or HIMS (see \citealt{Rev2000, Brocksopp2007, FHB09}), type C QPOs varying from 17 Hz to 31 Hz are observed in 
the PDS of observations during MJD 50968.87 to MJD 50975.6. 

\begin{figure}
\includegraphics[width=9cm]{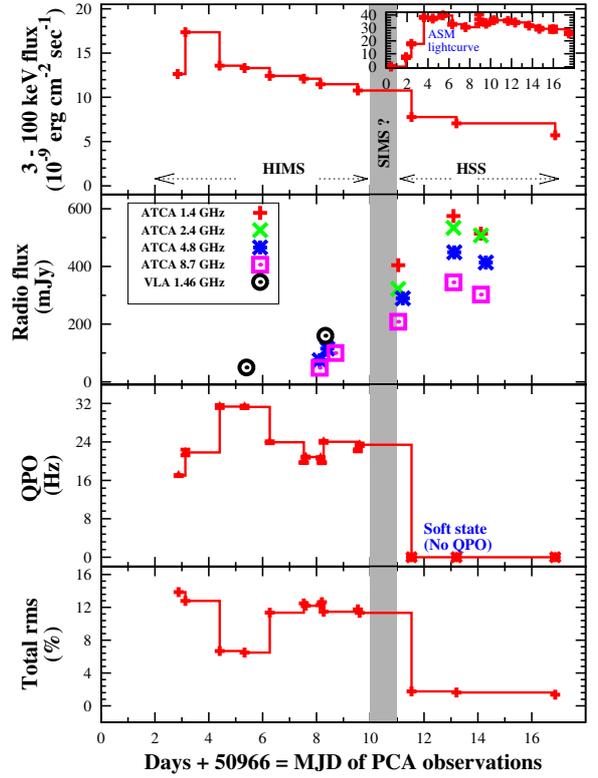}
\caption{Evolution of X-ray flux (3 - 100 keV), Radio flux, QPO frequency and total rms of the 
power spectra in 2 - 25 keV energy band during the 1998 outburst of XTE J1748$-$288; the X-ray 
spectral states and time of peak radio flare are marked. In the 3rd panel, we denote type C QPOs with triangles and observations where QPOs are not observed are shown using stars.}
\label{fig1-1748}
\end{figure}

For the next observation on MJD 50977.5, QPOs are not observed and total rms reduces to 1.8\%. We observe a drop in the 3 - 100 keV total flux and the hardness ratio reduces from 1.2 to 0.8. The fractional disk flux increases from 0.3 to 0.71 and fractional powerlaw flux decreases from 0.7 to 0.3 (Figure \ref{ratio-time}). These characteristics suggest that the source has reached a canonical HSS (see Table \ref{flares3}; \citealt{Rev2000,Brocksopp2007}). Hence in this source the transition to SIMS probably would have been very fast. Although the SIMS has not been observed by PCA and HEXTE, the ASM hardness ratio is found to decrease on MJD 50976.29 (Figure \ref{asm}, Table \ref{flares3}) during the shaded region in Figure \ref{fig1-1748}. \citealt{Brocksopp2007} reports an increase of radio flux to 404.5 mJy on MJD 50977.04 during the transition from hard to soft state, which attains a peak value of 574.7 mJy on 50979.09 (Figure \ref{fig1-1748}). 

\subsubsection{XTE J1550$-$564 (1998 outburst)}

\begin{figure}
\includegraphics[width=9cm]{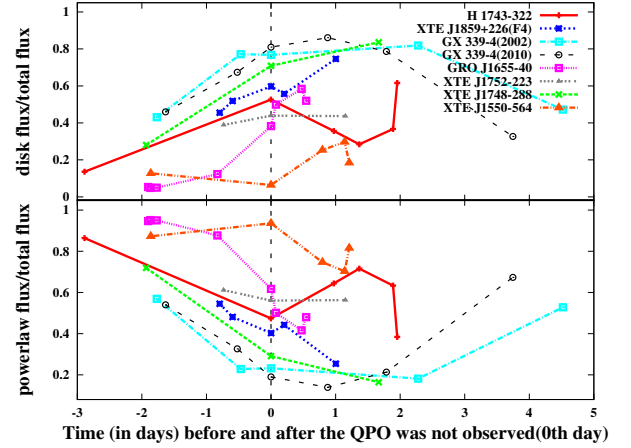}
\caption{Variation of fractional contribution of disk and powerlaw flux, during the period when for 
the first time the QPOs are not observed in the power spectra of the sources. The source MAXI J1836$-$194 which does not show presence of QPO during HIMS has been excluded from the plot.} 
\label{ratio-time}
\end{figure}

Type C QPOs are observed during the HIMS \citep{cbp09,Rao10}, while for the observation following this i.e. on MJD 51075.99 ($\sim$ 19 hrs later), the power spectra implies presence of two QPOs of 4.9 Hz (type B, Q-factor = 14.3, significance = 3.67, rms = 0.6\%) and of 13 Hz (Q-factor = 2.7, significance = 9.5, rms = 1.3\% indicates the QPO to be of type A), with significant reduction in the total rms of the PDS from 19.2\% to 2.56\% (see also \citealt{FHB09}). The 3 - 100 keV flux is observed to increase from 55 $\times$10$^{-9}$ erg cm$^{-2}$ sec$^{-1}$ to 
140 $\times$10$^{-9}$ erg cm$^{-2}$ sec$^{-1}$, and the contribution of disk flux also increases (Figure \ref{ratio-time}). Table \ref{flares3} indicates that the ASM hardness ratio also decreases to 0.85 on MJD 51076.69 (see \S\ref{Discu}). The radio flux is observed to increase from 
18 mJy to 168 mJy, and achieve a peak of 375 mJy on MJD 51078.26 \citep{Hanni2001}. It is possible that due to lack of observations, the transition to HSS is not indicated and the source is observed to enter HIMS.

\section*{} 
We do not observe rapid transition within SIMS or canonical transition to HSS for the source MAXI J1836$-$194 during its 2011 outburst, since it never occupied these spectral states. We observe that around the peak of the outburst, when the source is in HIMS during the period from MJD 55819 to MJD 55822, type C QPOs are present varying from 4 Hz to 5 Hz. Observation on MJD 55823.79 does not show the 
presence of QPOs in the PDS. The total rms of the PDS and the fractional 
contribution of disk flux remains constant around 0.1. Radio observations by RATAN-600 \citep{Trushkin2011} show an increase in radio flux to 50.02 mJy on MJD 55824.41, which has been reported to be corresponding to a steady/compact jet \citep{Russell2013}. We note that this is in similarity with the `spectro-temporal' characteristics observed during the radio flare peak of GX 339-4 in its 2002 outburst (see \S 3.1.1). Similar changes are observed during the first and second radio flares of V404 Cyg (see \S 3.1.3).

Tables \ref{flares}, \ref{flares2} and \ref{flares3}, summarize the details of variabilities observed during the jet ejection, in `spectro-temporal' properties. In these tables, `QPO F' is QPO frequency, `PCA HR' is hardness ratio obtained from PCA data, `D1' corresponds to duration (in days) of period when QPO is not observed and spectral softening (PCA and HEXTE) occurs after ejection, `D2' gives duration (in days) of SIMS observed, `ASM HR' is hardness ratio obtained from ASM one day averaged data. It has to be noted that the error in PCA hardness ratio is found to vary between 0.001 and 0.008, and the error in ASM hardness ratio varies between 0.03 and 0.08. Details of `spectro-temporal' characteristics observed in XTE J1859$+$226 have been earlier presented in \citealt{RN2014}, while that of V404 Cyg in \citealt{RAVS16}(arxiv:1601.03234). 

\begin{table*}
\caption{Details of `spectro-temporal' properties (in 2 - 25 keV band) of the observations during the flares/ejections in outbursting black hole sources. }
\label{flares}

\begin{tabular}{ccccccccc}
\hline 
MJD/ & Flare peak/ & QPO F  & QPO & Total & PCA HR & D1 & D2 & ASM HR\\
(PCA ObsID) &  ejection(mJy) & (Hz) & Type & rms(\%)  &  \\
\hline 
Source : GX 339$-$4 (2002 outburst) &\\

52406.49	& & & & & & & &	0.61\\
52406.70/ & & 5.77$^{+0.02}_{-0.04}$ & B & 0.15\\
(70110-01-11-00)&\\
52407.40	& & & & & & & &	0.59\\
52408.34	& & & & & & & &	0.49\\
 52408.48 & 10 - 55 \\
52409.388	& & & & & & & &	0.504\\
52410.360	& & & & & & & &	0.512\\
 52410.5/  & & 7.9${\pm0.2}$ & C & 9 & 0.4 \\
(70110-01-12-00)&\\
52411.33	& & & & & & & &	0.51\\
 52411.6/  & & 5.9${\pm0.05}$ & B & 5.5 & 0.25 & & $\sim$1 \\
(70109-01-07-00)&\\
 52412-52414  & & 0 & & 2.5 - 4.4 & 0.22 - 0.17 & $>$2  \\
52412.20	& & & & & & & &	0.45\\
52413.23	& & & & & & & &	0.42\\
 52413.42 & detection at 20.39 & \\
52414.26	& & & & & & & &	0.35\\
52415.72	& & & & & & & &	0.41\\
 52416.6/  & & 6.39${\pm0.04}$ & B & 9 & 0.29 & & $>$4 \\
(70110-01-14-00)\\

\hline
Source : GX 339-4 (2010 outburst) &\\

	 55303.6/ & & 5.71$^{+0.04}_{-0.03}$ & C &  14 & 0.4 \\
(95409-01-15-01) &\\
55304.25	& & & & & & & &	0.327\\
 55304.71/ & & 5.71$^{+0.02}_{-0.03}$ & B & 8.5 & 0.25 & & $>$1.5\\
(95409-01-15-02) &\\
55305.49	& & & & & & & &	0.33\\
    55305-55307/  & & 0 & & 2.8 - 4.3 & 0.14 - 0.2 & $>$3\\
(95409-01-15-03/04/05) &\\
 55303 - 55308 & reported \\
55306.41	& & & & & & & &	0.30\\
55307.92	& & & & & & & &	0.44\\
 55308.98/ & & 5.55${\pm0.03}$ & B & 6.8 & 0.27 & & $\sim$1\\
(95409-01-15-06)&\\
\hline
Source : XTE J1859$+$226 &\\

51464.28	& & & & & & & &	0.81\\
51466.07	& & & & & & & &	0.77\\

  51466.896/ & & 5.97${\pm 0.02}$ & C & 15.4 & 0.36 \\
  (40124-01-11-00) &\\
51467.31	& & & & & & & &	0.74\\

  51467.581/ & & 7.79$^{+0.3}_{-0.2}$ & A & 2.23 & 0.27 \\
  (40124-01-12-00)&\\
  51467.904 & F1 ejection at 5.3 & \\
  51467.961/ & & 6.1${\pm 0.02}$ & B & 3.78 & 0.27 & & $\sim$6  \\
  (40124-01-13-00)& \\
  51468.4 & F1 peak at 114 & \\
51468.51	& & & & & & & &	0.71\\\\
\cline{2-9}
51473.62	& & & & & & & &	0.75\\

  51473.89/ & & 7.67$^{+0.11}_{-0.09}$ & C* & 9.4 & 0.31 \\
  (40124-01-23-01)&\\
  51474.087/ & & 7.39${\pm 0.3}$ & C* & 7.82 & 0.29 \\
  (40124-01-15-02) &\\
  51474.287/ & & 0 & - & 6 & 0.25 & 1  \\
  (40124-01-15-03) &\\
51474.40	& & & & & & & &	0.73\\
  51474.429/ & & 5.87${\pm 0.07}$ & B & 6.28 & 0.29 & & $\sim$4 \\
  (40124-01-24-00)& \\
  51475 & F2 at 19 & \\
51475.68	& & & & & & & &	0.73\\

\hline

\end{tabular}
\\

\end{table*}

\begin{table*}
\caption{Continuation of the details on `spectro-temporal' properties (in 2 - 25 keV band) of the observations during the flares/ejections in outbursting black hole sources.}
\label{flares2}

\begin{tabular}{ccccccccc}
\hline 
MJD/ & Flare peak/ & QPO F & QPO & Total & PCA HR & D1 & D2 & ASM HR\\
(PCA ObsID) &  ejection(mJy) & (Hz) & Type & rms(\%)  &  \\
\hline 
Source : XTE J1859$+$226 &\\\\
51478.61	& & & & & & & &	0.74\\

  51478.980/ & & 7.50$^{+0.1}_{-0.05}$ & C* & 10.05 & 0.45\\
  (40124-01-31-00) & \\
51479.41	& & & & & & & &	0.67\\

  51479.635/ & & 0 & - & 1.8 & 0.25 & 4 \\
  (40124-01-32-00) &\\
  51479.940 & F3 at 31 & \\
51480.59	& & & & & & & &	0.64\\
51481.58	& & & & & & & &	0.62\\
51482.58	& & & & & & & &	0.62\\
  51483.106/ & & 4.87${\pm0.04}$ & B & 5.53 & 0.29 & & $>$ 1  \\
  (40124-01-36-00)&\\ 
\cline{2-9}
51483.72	& & & & & & & &	0.68\\
  51484.275/ & & 4.43${\pm0.02}$ & B & 3.78 & 0.43\\
  (40124-01-37-02) & \\
51484.62	& & & & & & & &	0.66\\
  51484.625 & F4 at 5.2 & \\
  51484.872/ & & 0 & - & 3.15 & 0.23 & 1\\
  (40124-01-38-00) &\\
51485.64	& & & & & & & &	0.54\\
  51485.874/ & & 6.97$^{+0.4}_{-0.2}$ & C* & 3.83 & 0.35 & & $\sim$ 3\\
 (40124-01-39-00) &\\
51486.64	& & & & & & & &	0.48\\
51487.63	& & & & & & & &	0.57\\
\cline{2-9}\\
  51487.009/ & & 4.77$^{+0.07}_{-0.08}$ & C* & 3.01 & 0.27 \\
 (40124-01-41-00) &\\
  51487.63 & F5 & \\
51487.63	& & & & & & & &	0.57\\
  51488.41/ & & 7.39$^{+1.1}_{-0.7}$ & C* & 3.92 & 0.31 \\
 (40124-01-43-00) &\\
  51488.48/ & & 7.34$^{+0.39}_{-0.34}$ & C* & 3.92 & 0.31 \\
 (40124-01-43-00)&\\
  51489.47/ & & 0 & - & 0.93 & 0.20 & 8\\
 (40124-01-49-00) &\\
51488.64	& & & & & & & &	0.498\\
51489.62	& & & & & & & &	0.482\\

\hline
Source : H 1743$-$322 &\\
54984.37/ & & 3.58${\pm 0.01}$ & C  & 22 & 1.36 \\
(94413-01-02-05) &\\\\
54985.47 & & & & & & & & 	1.03\\
54986.53 & & & & & & & &	0.87\\
54987.25 & & & & & & & &	0.71\\\\
 54987.26/   & & 0 & & 5.5 & 0.53 & & $>$2\\
(94413-01-03-00) &\\\\
	 54987.3 & ejection at 5.7 & - \\\\
	 54988 - 54989/ & & 0 & & 5 - 3.5 & 0.39 - 0.49  & 2 \\
(94413-01-03-01/07/05/06) \\\\
54988.61 & & & & & & & &	0.87\\
54989.51 & & & & & & & & 	0.31\\\\
	    54990.2 - 54990.4/ & & 3.7${\pm0.05}$ - 3.57${\pm0.2}$ & B - A & 5.3 - 3.4 & 0.54 & & $>$ 2 \\
(94413-01-03-02/03) &\\\\
54990.515 & & & & & & & &	0.73\\

\hline
\end{tabular}
\\

\end{table*}

\begin{table*}
\caption{Continuation of the details on `spectro-temporal' properties (in 2 - 25 keV band) of the observations during the flares/ejections in outbursting black hole sources. MAXI J1836$-$194 where transient jet was not observed (see \S \ref{results} and \S \ref{Discu}) is not included in this table. }
\label{flares3}

\begin{tabular}{ccccccccc}
\hline 
MJD/ & Flare peak/ & QPO F & QPO & Total & PCA HR & D1 & D2 & ASM HR\\
(PCA ObsID) &  ejection(mJy) & (Hz) & Type & rms(\%)  &  \\
\hline 

Source : XTE J1752$-$223 & \\
55217.536	& & & & & & & &	0.83\\
 55217.9 & compact at 20 &\\
55218.205	& & & & & & & &	0.55\\
	 55218.8/ & & 3.51${\pm0.02}$ & C & 8 & 0.29\\
(95360-01-01-00) &\\\\
	 55219.53/  & & 2.23${\pm0.2}$ & A & 7.36 & 0.25 \\
(95360-01-01-01) &\\\\
55219.54	& & & & & & & &	0.41\\
	 55220.1 & ejection at 6 \\
	 55220.68/  & & 2.16$^{+0.07}_{-0.06}$ & B & 6.7 & 0.25 & & $\sim$1.5\\
(95360-01-01-02) &\\\\
  55221 - 55281/ & & 0 & & $<$6 & $<$0.16 & 60 \\
(from 95360-01-01-03) &\\
55221.69	& & & & & & & &	0.19\\
55223.06	& & & & & & & &	0.17\\
	 55223 & peak at 18.8\\
\hline

Source : GRO J1655$-$40 &\\
 53440.68/ & & 6.37${\pm0.2}$ & B & 11 & 0.59 & & 2 \\
(91702-01-02-00) &\\\\
53440.74	& & & & & & & &	0.88\\
		 53440.8 & detection at 1.6 \\
   	       53441.5/  & & 16.37$^{+0.8}_{-0.2}$ & C & 3.8 & 0.35 \\
(91702-01-02-01) &\\\\
53441.73	& & & & & & & &	0.67\\\\
	  53442 - 53499/  & & 0 & & $<$5 & $\leq$0.3 & 57 \\
(from 91702-01-02-04) &\\\\
53442.88	& & & & & & & &	0.65\\
	 53449.8 & peak at 5.9 \\

\hline
Source : XTE J1748$-$288 &\\
 50975/ & & 23.4$^{+0.2}_{-0.3}$ & C & 14 & 1.2 \\
(30185-01-05-00) &\\\\
50975.57	& & & & & & & &	1.49\\
50976.29	& & & & & & & &	1.21\\\\

	 50977 - 51010/ & &  0 & & 1.5 & 0.8 & 33  \\
(from 30185-01-06-00) &\\\\
	 50977.4 & 404.5 \\
50977.65	& & & & & & & &	1.24\\

\hline

Source : XTE J1550$-$564 &\\
 51074.13/ & & 5.73$^{+0.02}_{-0.01}$ & C & 19.2 & 0.79 \\
(30191-01-01-00) &\\
51075.79	& & & & & & & &	0.95\\
		 51075.99/ & & 4.9${\pm0.04}$, 13${\pm0.1}$ & B, A(?) & 2.56 & 0.89 \\
(30191-01-02-00) &\\
51076.69	& & & & & & & &	0.85\\
		 51076.69 & ejection at 168 \\
		 51076.79/ & & 7.1${\pm0.02}$ & C &  9 & 0.76\\
(30191-01-04-00) &\\
51077.55	& & & & & & & &	0.81\\
51078.64	& & & & & & & &	0.86\\

\hline	
	
\end{tabular}
\\

\end{table*}

\begin{figure}
\includegraphics[width=9cm]{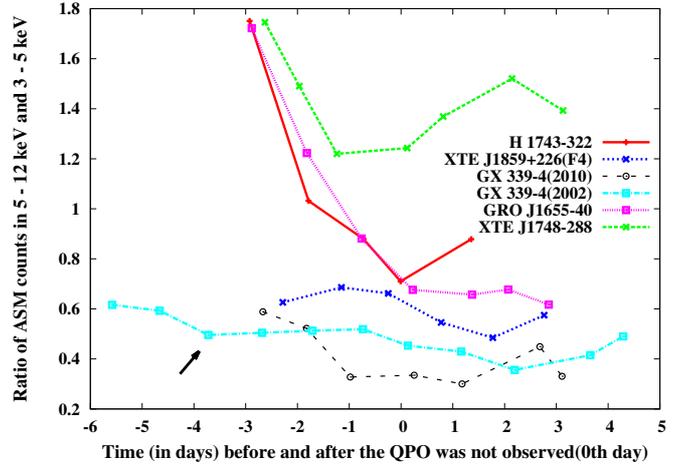}
\caption{Variation of ASM hardness ratio, during the period when for the first time the QPOs are not observed in the PCA power spectra of the sources. The arrow mark indicates the observation when the ASM hardness ratio decreases for GX 339$-$4 during the first radio flare of its 2002 outburst.} 
\label{asm}
\end{figure}

\section{Possible physical mechanisms}
\label{models}

Attempts have been made until now to explain the phenomena of jet ejection based on the rotation mechanism (spin) by 
\citealt{BZ1977,BP82,Meier2004,MG2004,dV2005} or due to the presence of poloidal magnetic field \citep{VarTag}. A unified model for the disk-jet coupling, has been arrived at by \citealt{FBG04} based on the observational results from several BH sources which arrives at a correlation between the X-ray characteristics and the jet ejection observed. They also suggest the formation of internal shocks in the steady jets during the hard state (see also \citealt{Drappeau2015}), and that its acceleration during the transition to intermediate state, powers the optically thin transient jets. \citealt{FHB09} suggests the possibility of the 
corona which gives rise to the hard photons, to be also responsible for the jet ejection. Recent studies suggest the accretion driven magnetic field to be responsible for the formation of transient jets \citep{Dexter2014}. But none of 
the above models have attempted to address in a single framework the fact that during an ejection event, QPOs are not observed in the power spectra which exhibits a decrease in total rms power, simultaneous softening of the energy spectra, faster timescales for the ejection event and subsequent hardening of the spectra.

The two component advective flow (TCAF) model developed by \citealt{ST95}, comprises of two types of flow: a Keplerian 
flow which has viscous timescale and a sub-Keplerian flow of free-fall timescale which surrounds the Keplerian disk and produces oscillating or standing shock fronts. The shocked flow heats up the post-shock 
region and a puffed up hot Compton cloud is formed, which intercepts the soft photons originating from the Keplerian disk 
and up-scatter them via inverse Comptonization to produce photons of higher energy. The oscillation 
of the shock front (i.e., the Comptonizing region) gives rise to QPOs (\citealt{dcnm14} and references therein). 
Recently, the TCAF model has been implemented into XSPEC as a table model (for 2.5 - 25 keV RXTE-PCA spectra by \citealt{Deb14}, and 0.5 - 100 keV for SWIFT-XRT, Integral-IBIS/RXTE-PCA spectra by \citealt{INM15}), and has been observed to successfully explain the 'spectro-temporal' characteristics, evolution of QPOs based on the Propagating Oscillatory Shock (POS) solution \citep{Ryu97,CM2000}, states occupied by the sources, the accretion flow parameters and the hard emission from the Comptonized corona (\citealt{Nandi2012, RN2014, INM15} and references therein). 

In order to understand the jet ejection phenomenon in outbursting BH sources, \citealt{Nandi2001} incorporated magnetic field into the TCAF. It is 
considered that the incoming matter due to accretion might have anchored large magnetic fields from the 
companion stars. During the LHS of the BH source these fields will be weaker. As the source 
rises towards the HIMS, the fields get amplified due to shear and the toroidal field will become stronger. During the 
transition to SIMS the flux tubes enter a hot region of temperature $>$ 10$^{10}$ K, and they collapse `catastrophically' before
entering into the Comptonized corona (see \citealt{CD94} and \citealt{Nandi2001} for details, and references therein). 
This results into the ejection of matter from the inner part of the disk 
(i.e. Compton cloud), which are observed as jets/flares in radio waveband based on the jet velocity. As a result, 
the oscillation of the corona ceases
and hence, QPOs will not be observed in the power spectra. Hence the 
energy spectra shall be dominated by thermal emission (i.e., disk component), or the high energy 
component (i.e., powerlaw flux) will be less, resulting in the softening of spectra 
(see also \citealt{SV2001,Chak2002,SV2003,RN2014}). Detailed modeling of how the ejected matter gets collimated resulting in relativistic motion have been presented in \citealt{Spru96,Fuk01,Chat04,Chat05}. In this paper, we attempt to interpret the disk-jet coupling 
observed in several BH sources within the context of magnetized TCAF model. The relativistic motion and collimation of the jet is not explored here since it is beyond the scope of this paper. 

\citealt{Motta2011} suggests that type A/C QPOs might be originating due to the Lense-Thirring precision of the inner hot flow while type B QPO occurs due to a sudden change of the surface density caused by the jet ejection. A correlation study performed between QPO types w.r.t the component fluxes suggests that type C QPOs are positively correlated with the disk flux (see \citealt{RN2014}). Also we note that the type A/B QPOs are observed after a jet ejection event. Hence, the argument that both type C and A QPOs have common origin need not be exactly correct.

It might be possible to understand about the origin of the different QPO types based on the two component advective flow model. During the rising phase of the outburst, the shock front propagates inward resulting the QPO frequency to increase while the size of the shock/shock location to decrease. Thus the frequency of the QPO is observed to be varying with time. Such an oscillation gives rise to type C QPOs, with increasing frequency during the LHS and HIMS of the rising phase. While in the declining phase the frequency of the QPOs decrease and since the rate of incoming matter reduces, the amplitude of QPO decreases.

After the jet ejection, the re-formation of Compton corona takes place and QPOs are observed again. We note that in most of the outbursting black hole sources the frequency value of QPO saturates while in the SIMS, and are observed to vary between 6 to 8 Hz. This implies that the propagation of shock towards the source has stalled for the duration of the SIMS. Ejection of matter from the corona has direct implication on shock front. Due to ejection of matter from the oscillating post-shock region, the shock front has been perturbed causing the shock strength to become weaker and subsequently the oscillating shock will produce different type of QPOs with weaker amplitude. This probably gives rise to A/B type QPOs, and in them having less amplitude in comparison to type C QPOs. Thus the origin of type A/B QPOs is probably related to the jet ejection event. 

A detailed modelling of the accretion dynamics during the jet ejection is required to have a better understanding on the origin of the type A/B QPOs. This study/investigation is beyond the scope of the present work which will be attempted later.

\section{Discussions}
\label{Discu}

In this paper, we have studied the `spectro-temporal' characteristics of several outbursting black hole 
sources, whenever a radio flare/jet emission occurs, in order to understand the disk-jet coupling and also the timescale of related events. 
We find that all the sources show that QPOs are not at all observed in the power spectra before the peak of radio flare (see also \citealt{Feroci99,SV2001,FHB09,MJ2012}). This suggests some correlation 
to exist between the absence of QPOs 
in power spectra and radio flare. We also observed that the total rms of the PDS reduces at the time 
of QPO suppression (see also \citealt{FHB09}). Along-with these characteristics, a spectral 
softening is implied by either an increase in contribution of disk flux or decrease in contribution 
of powerlaw flux for at least six sources (Figure \ref{ratio-time} and \ref{asm}). 
Hence as an after effect of the triggering of jet ejection, QPOs are not 
observed in the power spectra and `local' spectral softening is implied.

\subsection{Understanding the ejection for the sources studied based on the physical models}

The `spectro-temporal' characteristics of the source studied here follows the unified model of disk-jet coupling by \citealt{FBG04}. Apart from this, we find that the characteristics suggest more aspects during a jet ejection which are explained briefly in this subsection.
 
\subsubsection{`Local' spectral softening}

Based on the results in \S 3, we observe that for the sources H 1743$-$322, XTE J1859$+$226 and GX 339$-$4, a `local' softening occurs, within the soft intermediate state during the jet ejection. We find that this `local' softening is a very rapid event occurring within a few hours or days timescale, with the subsequent hardening occurring very fast. This can happen only if the matter is sub-Keplerian which flows in shorter timescale and is less viscous. This also suggests that the softening is most probably correlated with the jet ejection mechanism and may not be due to changes in Keplerian accretion rate. These characteristics suggest the magnetized TCAF model of \citealt{Nandi2001} to be applicable for the ejection event, whereas none of the other models (see \S\ref{models}), addresses such rapid events. 

The `local' softening is observed for GX 339$-$4, the multiple flares F2 to F5 of XTE J1859$+$226, V404 Cyg and H 1743$-$322. The duration of this softening is observed for $\geq$ 2 days (see Figure \ref{fig1-GX-2010}) in GX 339$-$4, V404 Cyg and H 1743$-$322, while it varies from 1 to 8 days in the case of XTE J1859$+$226 (see Table \ref{flares} and \ref{flares2}).  

We find that for the 2002 outburst of GX 339$-$4, although the PCA/HEXTE observations did not exist during the peak radio flare of 55 mJy on MJD 52408.48, the ASM observation on MJD 52408.38 and 52409.38 (see Table \ref{flares} and the arrow mark in Figure \ref{asm}) indicates the decrease in hardness ratio. This suggests of a possible spectral softening during the ejection event. 

The `local' softening observed in the 2002 and 2010 outburst of GX 339$-$4 during the transition to SIMS is observed to be repetitive. Similar features are observed during the multiple flares of V404 Cyg and XTE J1859$+$226 (Figure \ref{fig1-1859}; see also \citealt{RN2014, RAVS16}(arxiv:1601.03234)).

It has to be noted that during the 1st flare (F1) of XTE J1859$+$226, the softening is observed in 2 - 5 keV and 13 - 25 keV energy bands indicating a partial ejection, while during flares F2 to F5 the softening occurs in the entire 2 - 25 keV band suggesting that the corona has been disrupted completely. The complete ejection of the corona is probably not observed during F1, due to lack of X-ray observations. Our detailed analysis for these flares \citep{RN2014} verifies the `local' softening we observe in this paper for other BH sources. 

Our recent study of the source V404 Cyg during its June 2015 outburst, corroborates such `spectro-temporal' changes (see \S 3.7, and \citealt{RAVS16}(arxiv:1601.03234) for details), wherein we find similar `local' softening during the multiple flaring events. Here we observe the softening during an oscillating flare (see also \citealt{dcnm14}) in the transition from hard to intermediate state, and sudden flares while existing in the intermediate state itself.

\subsubsection{Canonical spectral softening}
We understand that for the sources GRO J1655$-$40, XTE J1752$-$223, XTE J1748$-$288 and XTE J1550$-$564, as an after effect of the jet ejection event the sources enter a very brief SIMS and then occupies a canonical soft state. Such rapid timescales for the occurrence of the ejection events and a short presence of SIMS can be understood based on the magnetized TCAF model which implies that the strength of the magnetic field increases during the transition to the SIMS ejecting out the matter in the Comptonized corona.

There is a lack of PCA and HEXTE data, during the first peak flare (a compact jet, see also \citealt{Brock13}) of XTE J1752$-$223. But an indication of spectral softening is suggested by the subsequent decrease observed in ASM hardness ratio (Table \ref{flares3}). In this source, we observe that just after the triggering of the jet ejection, partial absence of QPOs are observed while in the SIMS suggesting partial ejection of the Comptonized corona. This is in similarity with the X-ray characteristics during the 1st flare of XTE J1859$+$226 (see \S 5.1.1 and \citealt{RN2014}). 

During the 2005 outburst of GRO J1655$-$40 the radio flare achieves its peak after a time gap of $\sim$ 8 days since the initial increase. This probably suggests that although the ejection of Comptonized corona as jets is a transient event, the matter in the jet is perhaps moving very slow that it takes time to attain a peak flux.   

Although X-ray spectral softening is observed for the source XTE J1748$-$288, probably the transition to canonical HSS occurred through a SIMS (shaded region in Figure \ref{fig1-1748}). Although this SIMS has not been observed by PCA and HEXTE, a decrease in the observed ASM hardness ratio (Figure \ref{asm}, Table \ref{flares3}) during this period indicates the spectral softening.

Results of XTE J1550$-$564 show that the type B and A(?) QPOs are observed only in 5 - 20 keV. Although previous studies considered this observation to be part of a very high state or HIMS \citep{Sob2000}, the reduction of total rms (see also \citealt{FHB09}) and in ASM hardness ratio (see Table \ref{flares3}) implies this to be a SIMS. Here probably due to fast transition and lack of X-ray observations, a HSS could not be observed in PCA and HEXTE data. The ASM hardness ratio is observed to indicate a decrease implying a possible HSS (on MJD 51077.55), before the source is observed to enter a HIMS during the declining phase of the outburst.

\subsubsection{Caveats}
We find that for the source MAXI J1836$-$194, QPOs are not observed during the HIMS when a radio flare has been observed. Detection of radio flare during the HIMS of GX 339$-$4 (2002 
outburst) and the first and second flares in V404 Cyg, indicate similar characteristics. Since the source MAXI J1836$-$194 is not observed to have transited to a SIMS or HSS \citep{Ferrigno2012,Russell2014}, the system of jet ejection for such sources which have `failed outbursts' seems to be different as compared to other black hole sources. Although \citealt{Russell2013} suggests the presence of a compact jet, the fact that QPOs are not observed suggests of a probable transient jet ejection event.

\section*{}

A deeper look into the properties of the companions of these BHs suggests them to be of late F, G, K or M type (\citealt{San99,Hyn03}) which are in general magnetically active \citep{Linsky85}, with field strength around 1 kilogauss. This suggests that the models which specify the role of magnetic fields in formation of jets by evacuation of the Comptonized corona (see \S \ref{models} and \citealt{Nandi2001}) are applicable for these sources, wherein the magnetic field from the companion flows along-with the matter into the accretion disk. This field probably gets amplified during the rising phase of the outburst (i.e. the LHS and HIMS) and finally collapses to push out the matter in the corona \citep{FBG04, FHB09, Nandi2001} forming jets/flares, as the source enters the SIMS. 

The timescale required for amplification of the field is equivalent to the rising time of the source up-to the end of the HIMS when the jet ejection occurs. A typical black hole source with a fast rise and exponential decay light curve requires around 10 days to reach the transition point from HIMS to SIMS. For the sources studied in this paper which have a typical fast rise-time (XTE J1859$+$226, H 1743$-$322, XTE J1748$-$288, GRO J1655$-$40, XTE J1550$-$564), the timescale required for the field amplification is found to vary from 5 to 15 days. In these sources we find that the peak radio flux is higher and varies between 100 mJy and 600 mJy, except the 2009 outburst of H 1743$-$322 and 2005 outburst of GRO J1655$-$40 which have very less radio flux during the moment of ejection (5.7 mJy and 1.6 mJy respectively) and also while in the peak. For the sources which have a slow rise towards the X-ray peak require more time to reach the transition point i.e., they take larger timescale to amplify the magnetic field. As an example XTE J1752$-$223 requires 88 days, 48 days is taken by GX 339$-$4 during its 2002 outburst and 104 days for GX 339$-$4 during its 2010 outburst. For these sources the peak radio flux is found to be vary from 18 mJy to 55 mJy.

\citealt{FBG04} has performed detailed study on the estimation of the jet power based on the characteristics of the radio flare, like its rise time and flux density (see also \citealt{FM2016}). It is also possible to have a quantitative estimation of the kinetic jet power based on the X-ray characteristics of the source. Recently \citealt{ADN2015} have explored these by means of theoretical modeling (similar to the TCAF) and also estimated the jet luminosity for the source flux in its LHS and HIMS. This has been performed based on the X-ray flux estimated for a few of the sources quoted in this manuscript. The rate of mass outflow from the corona has been estimated to vary between 10\% to 18\% (see \S 5 of \citealt{ADN2015}). A detailed estimation of the same for the transient jets observed for the different sources, is beyond the scope of this paper.

The observational results for all the sources studied in this paper suggest the requirement of continuous and simultaneous Radio and X-ray observations, so as to observe exactly when the triggering of the jet ejection occurred. A similar concern has also been noted by \citealt{FHB09}. Although \citealt{MJ2012} could observe the moment of ejection in H 1743$-$322, they state that it is necessary to have dedicated observations in radio and X-ray, so as to find out if the pattern of the ejection event is similar for all the sources. For other black hole transient sources, till date there has been difficulty to find the moment of ejection and to have an exact correlation with the changes in their  `spectro-temporal' properties.

\section{Conclusions}
\label{Conc}

In this paper we have explored the variabilities in the `spectro-temporal' characteristics of several outbursting black hole sources. We have attempted to understand these within the context of the magnetized two component advective flow. We find that all the results for the several sources imply that the time duration between QPO not being observed and the jet ejection (Tables \ref{flares}, \ref{flares2} and \ref{flares3}) varies from hours to days, which indicates the possibility for the matter to be flowing not in viscous timescales, but in shorter timescales and the phenomenon to be occurring more rapidly. Hence these characteristics might be due to change in the accretion rate of the sub-Keplerian flow and not the Keplerian flow which has viscous timescales.

Thus based on the results of the `spectro-temporal' characteristics during the jet ejection for different sources, it can be summarized that

\begin{itemize}

\item Results for the sources GX 339$-$4, XTE J1859$+$226 (flares F2 to F5), V404 Cyg and H 1743$-$322 suggest that along-with canonical state transitions, just before the peak radio flare possibly there is a period during which the `spectro-temporal' characteristics are similar to the canonical HSS. This probably is a `local' HSS which might be occurring due to the jet ejection, and not the accretion rate.

\item The analysis of sources like XTE J1752$-$223, GRO J1655$-$40, XTE J1748$-$288 and XTE J1550$-$564 suggest the jet ejection to have occurred during the transition from HIMS to SIMS which exists only for a short duration or has been not observed due to lack of X-ray observations. These sources enter a canonical HSS after the ejection event. 

\item Results for the source MAXI J1836$-$194 suggest the ejection to have occurred probably during the HIMS itself although radio observations indicate presence of a compact jet. Observation of a radio flare during the HIMS of 2002 outburst of GX 339$-$4 also implies similar characteristics, which is evident from the decrease in ASM hardness ratio.

\item It might be possible in the near future to have a better understanding of the disk-jet coupling phenomenon based on continuous and simultaneous observations for different Galactic outbursting black holes with the help of ASTROSAT which has been launched recently, along-with GMRT or other radio observatories. 

\end{itemize}

\section{Acknowledgements}

We are grateful to Prof. A. R. Rao for his suggestions which helped a lot during the preparation of this manuscript. 
We are thankful to Prof. Richard Rothschild, Prof. Katja Pottschmidt and Prof. Gail Rohrbach for all 
the suggestions related to our queries on HEXTE data reduction; to Prof. 
Tomaso Belloni for the timely suggestions on the usage of `GHATS' package and other discussions. We thank Prof. Rob Fender, 
Prof. Catherine Brocksopp and Prof. James Miller-Jones for the information and suggestions 
related to the Radio observations.  

We are thankful to the reviewers whose valuable suggestions have helped in improving the manuscript. 

Radhika D., would like to thank Prof. B. R. S. Babu of University of Calicut for his guidance, 
and also acknowledges the research fellowship provided by ISRO Satellite Centre.
The authors also thank GD, SAG, for continuous support to carry out this research.

This research has made use of the data obtained through High Energy Astrophysics Science Archive 
Research Center on-line service, provided by NASA/GSFC.

\appendix
\section{Evolution of power density spectrum during jet ejection}
Here, we present the figures of the power density spectral evolution during the period a jet ejection/radio flare has occurred. We have included only those sources which exhibit a `local softening (local HSS)'.

Figure A1 shows the power spectral evolution of GX 339$-$4 during its 2010 outburst. A type C QPO is observed during MJD 55303.6 while the source is in the hard intermediate state, and a type B QPO exists during MJD 55304.71. The following observations during MJD 55305, 55306 and 55307 do not show any signature of QPOs. Radio detection has been reported during this period by \citealt{Russell2010} and \citealt{CadolleBel2011}. The next observation during MJD 55308.98 indicates presence of a type B QPO. We understand that the period of MJD 55305 to 55307 depicts `local softening'. A detailed discussion of this has been given in \S 3.1.2, 5.1.1 and Figure \ref{fig1-GX-2010} of the manuscript. 

\begin{figure}
\centering
\includegraphics[width=9cm]{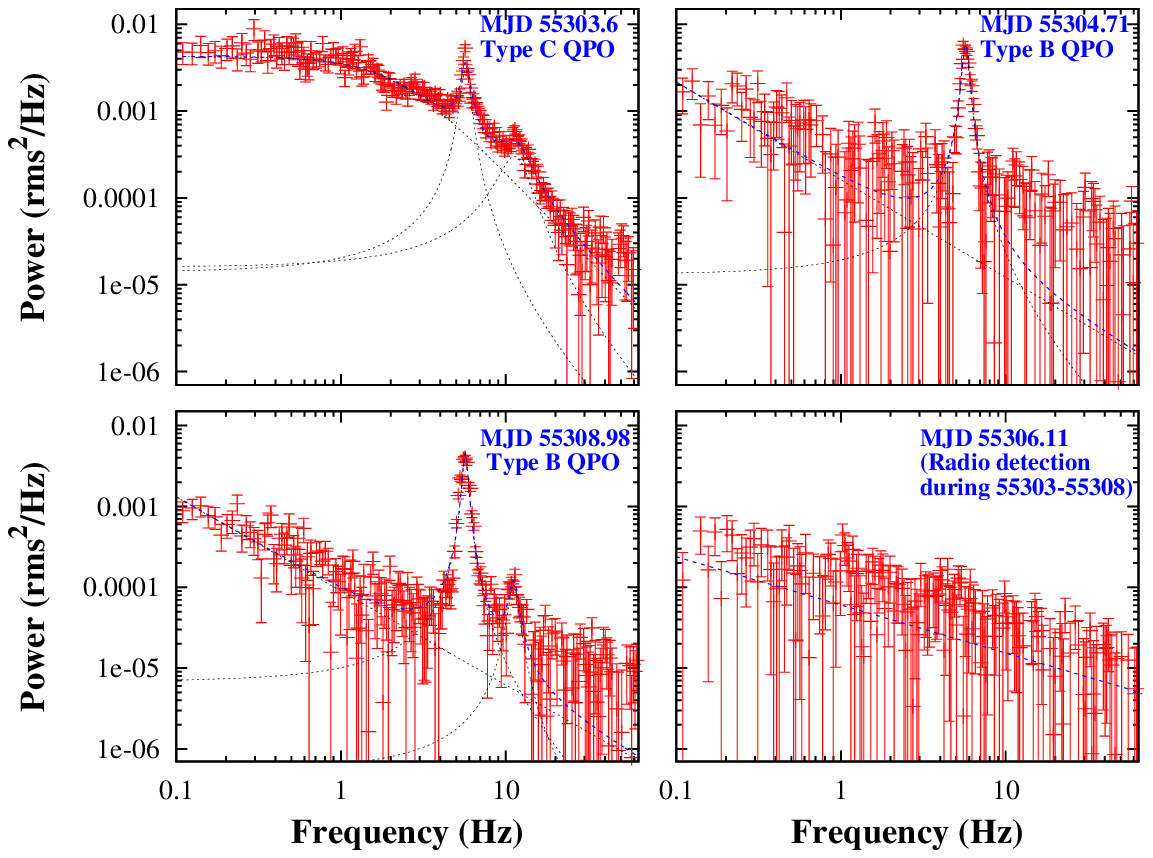}
\caption{Evolution of power spectra in clock-wise direction, during the 2010 outburst of GX 339$-$4.}
\end{figure}

In Figure A2, we show the evolution of power density spectra during the flare F3 of XTE J1859$+$226. We observe a type C* QPO during MJD 51478.980, while for the next observations (i.e. MJD 51479 to 51482) we do not observe QPOs. The flare F3 is observed during MJD 51479.90. Type B QPO is observed on MJD 51483.106 which is 4 days after the radio peak. The period of MJD 51479 to 51482 corresponds to `local softening' (see \S 3.1.2, 5.1.1. and Figure \ref{fig1-1859} of the manuscript).

\begin{figure}
\centering
\includegraphics[width=9cm]{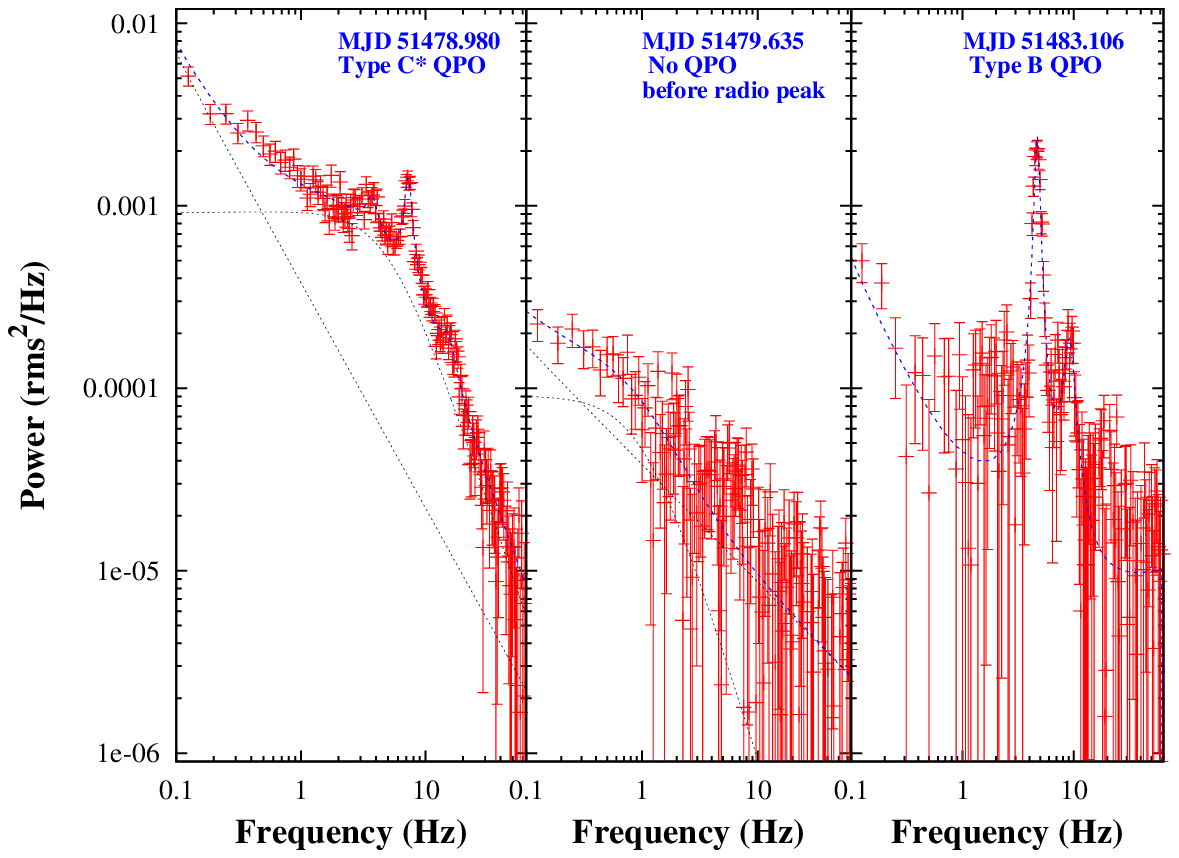}
\caption{Power spectral evolution during the third flare (F3) of XTE J1859$+$226, during its 1999 outburst.}
\end{figure}

Figure A3 shows the power spectral evolution of H 1743$-$322, indicating presence of type C QPO during the hard intermediate state on MJD 54984.37. Subsequent observation during MJD 54987.26 do not show presence of a significant QPO. \citealt{MJ2012} has reported the detection of a jet ejection event during MJD 54987.3. Signature of weak QPOs/peaked component is observed during MJD 54988.62 and 54989.22. Type B QPO is observed during MJD 54990.26. The period of MJD 54985 to 54989 belongs to `local softening' (see section 3.1.4, 5.1.1 and Figure \ref{fig1-H} in the manuscript). 

\begin{figure}
\centering
\includegraphics[width=9cm]{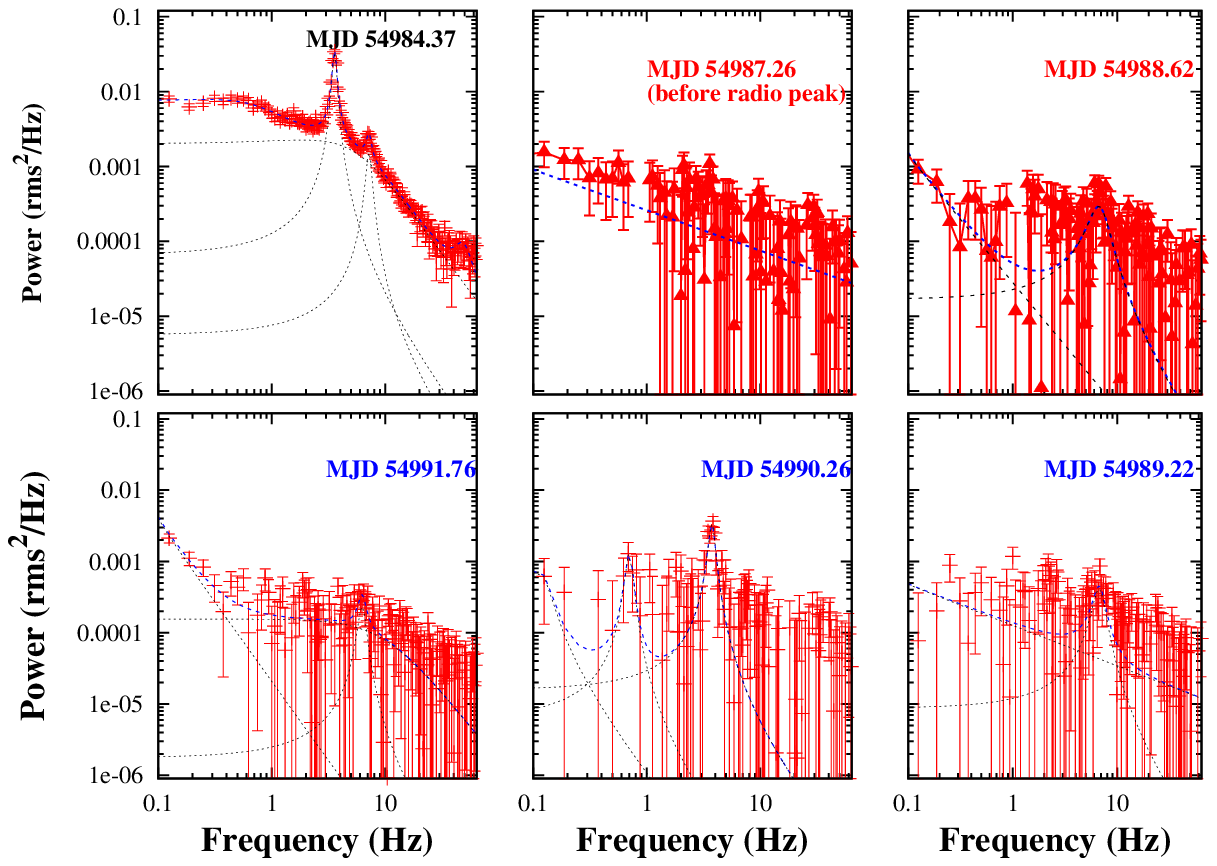}
\caption{Evolution of power density spectra in clock-wise direction during the 2012 outburst of H 1743$-$322.}
\end{figure}

\bsp

\label{lastpage}

\end{document}